# Thermal transport in MoS$_2$/Graphene hybrid nanosheets

Zhongwei Zhang[1], Yuee Xie[1†], Qing Peng[2], and Yuanping Chen[1*]

[1] Department of Physics, Xiangtan University, Xiangtan 411105, Hunan, P.R. China

[2] Department of Mechanical, Aerospace and Nuclear Engineering, Rensselaer Polytechnic Institute, Troy, NY, 12180, USA

**Abstract:** Heat dissipation is a very critical problem for designing nano-functional devices, including MoS$_2$/Graphene heterojunctions. In this paper we investigate thermal transport in MoS$_2$/Graphene hybrid nanosheets under various heating conditions, by using molecular dynamics simulation. Diverse transport processes and characteristics, depending on the conducting layers, are found in these structures. The thermal conductivities can be tuned by interlayer coupling, environment temperature and interlayer overlap. The highest thermal conductivity at room temperature is achieved as more than 5 times of that of single layer MoS$_2$ when both layers are heated and 100% overlapped. Different transport mechanisms in the hybrid nanosheets are explained by phonon density of states, temperature distribution, and ITR. Our results not only could provide clues to master the heat transport in functional devices based on MoS$_2$/Graphene heterojunctions, but also are useful to analyze thermal transport in other van der Waals hybrid nanosheets.

## I. Introduction

The intrinsic zero band gap of graphene limits its extensive applications in electronics and optoelectronics [1,2], although the first single-layer nanosheet exhibits many outstanding properties such as super-high carrier mobility and super-high thermal conductivity [3-6], in addition to its superior stiffness [7]. After graphene, many other nanosheets were proposed and synthesized successfully in recent years [8-11]. In all of the new nanosheets, single-layer molybdenum disulfide ($MoS_2$) has attracted much attention because of its direct band gap and high carrier mobility [12-14]. A lot of theoretical studies have been carried out to explore the electronic and optoelectronic properties of $MoS_2$ nanosheets [14-17] as well as mechanical properties[18]. Experimentally, single-layer $MoS_2$ transistors have been reported, which show high performance, such as a mobility of at least 200 $cm^2V^{-1}S^{-1}$ and room-temperature current on/off ratios of $1\times10^8$ [14,15]. These studies indicate that the single-layer $MoS_2$ may be an ideal electronic and optoelectronic material. However, the thermal conductivity of $MoS_2$ is much lower, which may induce problems in its thermal stability and restrict development of devices based on the new nanosheet [19-21].

In order to combine the advantages of single-layer $MoS_2$ and graphene nanosheets, various functional devices based on $MoS_2$/Graphene hybrid nanosheets (MGHN) have been a new focus recently [22-27]. For example, Chang *et al.* studied a MGHN with external electrodes applied on both layers, as shown in Fig. 1(a) [23]. High electrochemical performances for lithium ion batteries are exhibited in these layered composites. Zhang *et al.* reported that another hybrid nanosheet in Fig. 1(b), where the $MoS_2$ layer is a transport layer while graphene is a substrate, is a potential photodetector with high photoresponsivity [24]. A field-effect transistor based on a MoS$_2$/Graphene heterojunction is reported by Kwak *et al.* and Myoung *et al.*, which has a large current modulation, spin-dependent tunneling, and lower barrier height [25,26]. The heterojunction is illustrated in Fig. 1(c). Although these functional devices based on MGHN show excellent electronic or optoelectronic properties, to the best of our knowledge, thermal transports in these hybrid nanosheets have never been reported. It is natural to ask how about the ability of these devices to conduct heat and what new transport phenomena these structures have.

In this paper, thermal transports in three types of MGHNs, as shown in Fig. 1, are studied by using molecular dynamics methods. Although all of the three structures consist of the same MoS$_2$ and graphene layers, thermal transport processes within are completely different, as well as their abilities to conduct heat. The first MGHN (MGHN-1) in Fig. 1(a) has the highest thermal conductivity, followed by MGHN-2 in Fig. 1(b), while the thermal conductivity of MGHN-3 in Fig. 1(c) is the lowest. We analyze the transport mechanisms in these hybrid nanosheets by using the phonon density of states (PDOS), temperature distribution, and interlayer thermal resistance (ITR). It is interesting to find that ITR has a positive relation to thermal transport in MGHN-1, while the transport in MGHN-2 and MGHN-3 are negative. In addition, the relation between interlayer coupling strength, environment temperature, number of graphene layers, and thermal conductivities of these structures is discussed.

## II. Model and Simulation method

Three types of MGHNs were considered as shown in Fig. 1. In all structures, MoS$_2$ and graphene layers have the same width and length, labeled by *W* and *L*, respectively. Figure 1(a) shows the side view of MGHN-1, where MoS$_2$ and graphene layers are entirely coupled together, and heat source and heat sink are applied on both the MoS$_2$ and graphene layers. The hybrid nanosheets in Fig. 1(b) are MGHN-2, where the heat source and heat sink are only applied to the MoS$_2$ layer. Therefore, the graphene layer in this structure is somewhat like a substrate. The top view of MGHN-1 or MGHN-2 is shown in Fig. 1(d). Due to different lattice constants, there exists lattice mismatch between graphene and MoS$_2$ layers. Here, the unit cell of the hybrid structure is constructed by matching a 4×4 supercell of MoS$_2$ and a 5×5 supercell of graphene with a tensile strain of ~1.5% (see the diamond box). The lattice constant of the unit cell is $a$ = 12.48 Å, while the lateral width of the unit cell is $b$ ($=\sqrt{3}/2a$). Then, the width $W$ and length $L$ of the hybrid structure can be defined in units of $a$ and $b$, respectively, and we set $W = 6b$ and $L = 12a$. Figure 1(c) shows the side view of MGHN-3, which is a MoS$_2$/Graphene heterojunction by the overlap of partial MoS$_2$ and graphene layers. The length of the overlap region is labeled by $L'$, and the stacking form of the overlap region is the same to the former two structures. The heat source and heat sink are separately applied on one side of MoS$_2$ and graphene layers. Besides these three bilayer structures, we also consider trilayer structures that another graphene layer is covered on MGHN-1 or MGHN-2, i.e., Graphene/MoS$_2$/Graphene nanosheets (GMGN). The sandwiched structures are named GMGN-1 and GMGN-2 corresponding to MGHN-1 and MGHN-2, respectively (not shown).

The simulations are carried out by using the large-scale atomic/molecular massively parallel simulator (*LAMMPS*) package [28]. To model the interactions of C-C atoms, the Adaptive Intermolecular Reactive Empirical Bond Order (AIREBO) potential is adopted [29]. A recently developed Stillinger-Weber (SW) potential is used to describe the interactions in MoS$_2$ layer [21]. In all the structures in Fig. 1, the MoS$_2$ layer is coupled to the graphene layer through a *van der Waals* force, i.e., the interlayer coupling originates from the *van der Waals* force. To model the weak *VDW* forces, the 12-6 Lennard-Jones (LJ) potential is adopted as follows [30]:

$$E = 4\eta\varepsilon\left[\left(\frac{\sigma}{r}\right)^{12} - \left(\frac{\sigma}{r}\right)^{6}\right] \quad (1)$$

here, $r$ represents the distance of two atoms, $\varepsilon$ is the energy that reflects their interaction strength, $\sigma$ denotes the zero-across distance of the potential, and $\eta$ is a scaling factor and can be used to tune the interaction strength ($\eta = 1$ by default). The original LJ potential parameters for Mo-Mo, S-S and C-C atoms are taken from Ref. [31-33]. Determined by arithmetic and geometric mixing rules [34], the potential parameters for Mo-C atoms are set as $\varepsilon = 44.758$ meV and $\sigma = 2.949$ Å; for S-C atoms are set as $\varepsilon = 12.035$ meV and $\sigma = 3.390$ Å, respectively. The *VDWs* interaction cutoff distance is chosen to be 10 Å. The hybrid structures are optimized under NPT condition. After optimization, the lattice constant of the unit cell is $a = 12.43$ Å, corresponding to a stress strain of ~0.35% in $MoS_2$ and a tensile strain of ~1.1% in graphene. The width and length of all graphene and $MoS_2$ layers are $W = 6.46$ nm and $L = 14.93$ nm, respectively. The interlayer distance between $MoS_2$ and graphene is calculated as 3.35 Å, which coincided well with the reported parameter of 3.32 Å [35].

In our simulations, the hybrid structures are initially relaxed in Canonical ensemble (NVT) conditions for 50 picoseconds (*ps*), with a time step of 0.46 femtoseconds. Their left/right boundaries are fixed while the other two are periodic boundaries. Next to the fixed boundaries the adjacent two cells are coupled to Nosé–Hoover thermostats, to achieve temperature gradients, with temperatures $T_0+\Delta T$ and $T_0-\Delta T$, respectively. Herein, $T_0$ is the environment temperature and the temperature drop is $\Delta T = 10\% \ T_0$. From Fourier's law, the thermal conductivity $K$ is defined as [36]:

$$K = \frac{J}{\nabla T \cdot S} \tag{2}$$

where $\nabla T$ is the temperature gradient and $J$ is the heat flux from the heat source to heat sink which can be obtained via calculating the heat baths power. $S=W\times H$ is the cross-sectional area, $W$ is the width, and $H$ is the thickness of the hybrid structure, which we have chosen for $MoS_2$ to be 3.66 Å [27] and 1.42 Å for graphene [37,38]. To reach the non-equilibrium steady state, the system was relaxed for 60 *ps* under NVE ensemble. After 50 *ps*, the temperature difference has reached the steady state. During this procedure, the total energy and average temperature both were fluctuating around targeted values. After then, 100 *ps* was used to calculate the thermal conductivity. In all of our simulations, the hybrid structures show good stability even in high temperature, by checking their atomic configurations and redial distribution functions (not shown). It indicates that the potential we selected is effective and reliable.

To understand the underlying mechanisms of phonon transport, the phonon density of states (PDOS) has been studied. The PDOS is calculated from the Fourier transform of the velocity autocorrelation function [39]:

$$\text{PDOS}(\omega) = \frac{1}{\sqrt{2\pi}} \int e^{-i\omega t} \langle \sum_{j=1}^{N} v_j(t) v_j(0) \rangle \tag{3}$$

where $v_j(0)$ is the average velocity vector of a particle $j$ at initial time, $v_j(t)$ is its velocity at time $t$, and $\omega$ is the vibration wave number.

### III. Results and discussion

We firstly study the homogeneous temperature gradient in both layers as in MGHN-1 structure. The thermal conductivity of MGHN-1 as a function of environment temperature $T_0$ is calculated as shown in Fig. 2. To calculate thermal conductivity of MGHN-1, the total system cross-section is included. For comparison, thermal conductivities of single-layer $MoS_2$, and graphene are also presented. One can find that the thermal conductivity of graphene increases while that of $MoS_2$ decreases as $T_0$ ascends; moreover, the former is much larger than the latter especially at high temperature. The conductivity of MGHN-1 is between them, which is about 5 and 1/3 times that of $MoS_2$ and graphene, respectively, at room temperature (300K). The coupling between layers might play an important role in the thermal transport in the hybrid structures. To examine this interlayer-coupling effect, we compute the thermal conductivity of MGHN-1 without interlayer coupling, as shown in Fig 2.

The comparison between the conductivities of MGHN-1 with and without interlayer coupling indicates that the interlayer coupling reduces the thermal conductivity (see the red solid and blue dotted lines). The higher the temperature, the more significant reduction of the conductivity. This implies that the interlayer coupling is enhanced with respect to an increment of temperature. As a further study, we investigate the sandwiched nanosheets GMGN-1. We find that the thermal conductivity of the sandwiched nanosheets GMGN-1 is higher than the thermal conductivity of MGHN-1, but lower than that of MGHN-1 without interlayer coupling after $T_0$ is higher than room temperature (see the green line in Fig. 2). This further indicates that the coupling between $MoS_2$ and graphene layers has a large effect on the thermal transport in MGHN-1.

The mechanism of thermal transport in MGHN-1 is schematically displayed in Fig. 3(a). As the heat source and heat sink are applied on both $MoS_2$ and graphene layers, the temperature gradient are homogeneous in out-of plane. In other words, there is no temperature drop in the vertical direction, and thus no net heat flux exists between two layers. The heat fluxes only flow in the $MoS_2$ or graphene layer, as shown by the red arrows. However, the interlayer coupling will lift the interlayer exchange of phonons and strengthen interface scattering [34,40]. We compare the out-of plane PDOS for isolated graphene and graphene after it is coupled with $MoS_2$ in Fig. 3(b). It is seen that the PDOS is reduced by the interlayer coupling, especially for the phonons whose frequencies are less than 15 THz. The phonon frequency range of single-layer $MoS_2$ is only from 0 to 15 THz [41,42]. This demonstrates that all phonon modes in the $MoS_2$ layer have interactions with those in the graphene. The interaction of phonon modes enhances the phonon scattering, as a result the thermal conductivity of the hybrid nanosheets decreases. The interactions of phonon modes are also tightly associated with the environment temperature $T_0$. With an increase in $T_0$, high-frequency phonons gradually participate in the transport process, and thus the phonon interactions between the two layers become more dramatic. This is the reason why at higher temperatures the differences in thermal conductivities between MGHN-1 with and without coupling are larger.

The interlayer interactions of phonons can also be reflected by ITR between two layers, which is usually expressed as: $R = \Delta T/J$, i.e., temperature change $\Delta T$ divided by heat flux $J$. To calculate the ITR between $MoS_2$ and graphene layers, we set the graphene and $MoS_2$ layers as heat source and sink, respectively, and the temperature drop is $\Delta T = 10\% \ T_0$. Figure 3(c) shows the ITR between $MoS_2$ and graphene as a function of temperature. It is seen that the ITR decreases with an increase in temperature, indicating that the phonons are more easily transmitted from one layer to another. The phonon communication between the two layers enhances the interlayer scattering, and reduces the thermal conductivity. Therefore, in MGHN-1, the thermal conductivity is in direct proportion to ITR.

The thermal conductivity of MGHN-2 as a function of environment temperature $T_0$ is shown in Fig. 4(a), where the blue and red lines represent the cases for MGHN-2 and GMGN-2, respectively. The geometry of MGHN-2 is similar to that of MGHN-1. However, the thermal function of the graphene layer in MGHN-2 is something like a substrate because the heat baths are only applied on the $MoS_2$ sheet in MGHN-2. Therefore, when we calculate the thermal conductivity of the structure by using Eq. (2), the cross-sectional area $S$ only includes the cross-section of $MoS_2$. One can find from Fig. 4(a) that the thermal transport of the $MoS_2$ sheet is improved by the help of the graphene substrate(s), comparing with the thermal conductivity of isolated $MoS_2$ in Fig. 2. The thermal conductivities of MGHN-2 and GMGN-2 are about 2.0 and 2.6 times that of the isolated $MoS_2$ at $T_0 = 300K$, respectively. The dotted lines in Fig. 4(a) present increased thermal conductivities as a function of temperature under the help of graphene sheet(s). The trend indicates that the contributions of the graphene layers increase with $T_0$.

To investigate the mechanisms of thermal transport in MGHN-2, the temperature spatial distributions are calculated, as shown in Fig. 5(a). The black and blue lines present variations of temperature along the transport direction in the $MoS_2$ and graphene layers, respectively. For comparison, variation of temperature in an isolated $MoS_2$ is also shown. One can see that the temperature of the isolated $MoS_2$ drops linearly along the transport direction because the heat energy is conservative in the MoS$_2$ sheet. As the MoS$_2$ layer is coupled to a graphene substrate, its temperature drop is no longer linear, implying that the heat energy in the MoS$_2$ layer is not a constant anymore because some heat flow to the substrate. In Fig. 5(a), the temperature of the graphene layer seems to be a constant, but a close look in the inset shows that it also drops nonlinearly even if the temperature drop is very small. The temperature difference between the two contacted layers indicates that there exists heat exchange between them.

We use Fig. 5(b) to illustrate the process of thermal transport and exchange in MGHN-2. At the left high-temperature side, heat fluxes flow vertically from MoS$_2$ to graphene because the temperature of the MoS$_2$ layer is higher than that of graphene. The quantity of the vertical flux decreases along the horizontal direction due to the decrease of the temperature difference. The heat energy obtained from the MoS$_2$ layer flows from left to right sides in the graphene layer, and then flows back to the MoS$_2$ layer, because at the right side the temperate of graphene is higher than that of MoS$_2$. In the horizontal direction, there are two thermal transport channels, one is in the MoS$_2$ layer and the other is in graphene. Because the graphene is a high-conductivity material, a lot of heat will transport across it as its two sides have a small temperature drop. Therefore, although the phonon scattering induced by vertical coupling will decrease the intrinsic thermal conductivity of MoS$_2$, the opening of the additional channel in the graphene substrate improves the ability of the whole device to conduct heat.

As discussed above, the ITR between MoS$_2$ and graphene layers decreases with the increase of environment temperature $T_0$ (see Fig. 3(b)). Therefore, at high $T_0$, thermal exchange between the two layers increases, i.e., the heat energy carried by the graphene layer increases. This explains the reason why the thermal conductivity in Fig. 4(a) increases with $T_0$. In this case, the thermal conductivity is inversely proportional to the ITR.

To fine tune the interlayer coupling, we introduce the variable $\eta$ as coupling strength to scale the interfacial couplings between layers as expressed in Equation 1. We find that the thermal transport in the MGHN-2 structure is dependent on not only the environment temperature but also the interfacial coupling strength $\eta$. As shown in Fig. 4(b), the thermal conductivities increase with the coupling strength, which can also be attributed to the decrease of ITR (see the black line in Fig. 4(b)). When $\eta = 2.0$, the thermal conductivities is about 1.4 times that of $\eta = 1.0$. So, one can improve the thermal transport in MGHN-2 by applying a vertical compressive strain.

In Fig. 6(a), the thermal conductivity of MGHN-3 as a function of $R_{\text{overlap}}$ ($L'/L \times 100\%$) is shown at $T_0 = 300$ K. In this structure, the cross-section of $MoS_2$ is included to calculate thermal conductivity. One can find that thermal conductivity increases with the overlap ratio. At 50% overlap ratio, i.e., half contact of two layers, the thermal conductivity is 3.4 W/mK, which is close to the value of isolated $MoS_2$, while the thermal conductivity of full contact is about 4.5 W/mK. Figure 6(b) shows the modulation of interfacial coupling strength $\eta$ on thermal conductivity for the MGHN-3 structure with 50% overlap ratio. The ITR between the two layers decreases with the coupling strength and the environment temperature $T_0$. As a result, the thermal conductivities increase with $\eta$ and $T_0$. This is similar to the case of the MGHN-2 structure. However, the thermal conductivities of the MGHN-3 structure are lower than those of MGHN-1 and MGHN-2.

To explore the thermal transport process in MGHN-3, we calculate the temperature distributions of MoS$_2$ and graphene layers, as shown in Fig. 7(a). Here, the two layers are half contact (50% overlap ratio) and the temperatures of the left and right sides are 330 K and 270 K, respectively. It is seen that the temperature drop in the graphene layer is very small, while a temperature drop $\Delta T \approx 30$ K is observed in the MoS$_2$ layer. According to the temperature distributions, Fig. 7(b) presents the distribution of heat fluxes in the structure. The heat flux horizontally flows in the graphene layer from left to right, at the overlap region it vertically flows from graphene to MoS$_2$ layers because of the temperature difference of two layers, and then horizontally flows in the MoS$_2$ layer from left to right. In the whole process, the heat energy should be conservative. Because the thermal conductivity of graphene is much larger than that of MoS$_2$, the temperature drop of graphene layer is very small while the MoS$_2$ layer has a big temperature drop.

In the MGHN-3 structure, the overlap region is the transport bottle neck, because the ITR (interlayer thermal resistance) is higher than the intralayer thermal resistance. Therefore, the thermal conductivity of MGHN-3 structure can be enhanced by weakening the ITR, such as increasing the overlap ratio and increasing the coupling strength.

It should be noted that in real applications most devices are supported on a substrate of insulating layer, such as SiO$_2$. The insulator layer should have some effect on the thermal transport in the devices. The previous studies showed that, as a graphene is supported by an insulator, thermal conductivity of the graphene drops slightly because of the phonon scattering between layers [43,44]. Therefore, as MGHN-1, MGHN-2 and MGHN-3 are placed on a substrate of insulator layer, their thermal conductivities would also decrease slightly.

## IV. Conclusions

In summary, we have studied thermal transport in three types of hybrid $MoS_2$ and graphene structures with three heating conditions, by using molecular dynamics simulations. These structures show diverse transport processes and abilities of thermal transport. The MGHN-1 structure has the highest thermal conductivity (about 5 times of $MoS_2$), because the graphene layer can carry most of the heat energy. Although the interlayer scattering induced by interlayer coupling reduces the out-of plane PDOS, the super-higher conductivity of the graphene layer makes the structure a good thermal transport. In the MGHN-2 structure, the $MoS_2$ layer is the main layer to transport heat while the graphene layer is just a substrate. By a process of heat transfer between $MoS_2$ and graphene layers, graphene can only transport a small part of heat, and thus its thermal conductivity is lower than that of the MGHN-1 structure (about 2~3 times of $MoS_2$). The thermal transport ability in the MGHN-3 structure is the lowest, because there is a bottle neck of transport at the contact region between the two layers. In addition, the conductivities of these structures can be slightly tuned by the interlayer coupling strength, environment temperature, and contact area. These findings could improve our knowledge of $MoS_2$ based hybrid structures that may be useful for the applications of $MoS_2$ in nanoscale devices.

**Acknowledgments**

This work was supported by the National Natural Science Foundation of China (Nos. 51176161, 51376005 and, 11474243).

Corresponding author: † xieyech@xtu.edu.cn; * chenyp@xtu.edu.cn

**Figure captions**

**Figure 1.** Three types of MGHNs. (a) MGHN-1, where heat source and heat sink are applied on both the $MoS_2$ and graphene layers. (b) MGHN-2, where heat source and heat sink are only applied on the $MoS_2$ layer. (c) MGHN-3, where the $MoS_2$ and graphene layers are connected by an overlap area and heat source is applied on one side of graphene and heat sink on $MoS_2$ layers, respectively. (d) A top view of MGHN-1 and MGHN-2. The green diamond box is the unit cell of the hybrid structure, which is constructed by matching a 4×4 supercell of $MoS_2$ and a 5×5 supercell of graphene. The lattice constant of the unit cell is $a = 12.43$ Å, while the lateral width of the unit cell is $b$ $(=\sqrt{3}/2a)$. In all structures, the width and length of the $MoS_2$ and graphene nanosheets are $W = 6.46$ nm and $L = 14.93$ nm, respectively. Width and length of the overlap region in MGHN-3 is $W$ and $L'$, respectively.

**Figure 2.** Thermal conductivities of MGHN-1, MGHN-1 without interlayer coupling, GMGN-1, single-layer $MoS_2$, and graphene as functions of temperature $T_0$.

**Figure 3.** (a) Schematic view of the thermal transport process in MGHN-1. Red arrows show the heat fluxes in the two layers while black arrows illustrate exchange and scattering of phonons. The color bar labels the temperature scale in the structure. (b) The out-of plane PDOS for isolated graphene and coupled graphene in the MGHN-1. (c) ITR between $MoS_2$ and graphene layers as a function of temperature $T_0$.

**Figure 4.** (a) Thermal conductivities of MGHN-2 and GMGN-2 as functions of temperature $T_0$ (solid line), and the increased thermal conductivity, compared to isolated ones, as functions of temperature $T_0$ (dotted line). (b) Thermal conductivity

of MGHN-2 (blue line) and ITR between MoS$_2$ and graphene (black line) as functions of coupling strength η, at $T_0$=300 K.

**Figure 5.** (a) Horizontal temperature distribution of isolated MoS$_2$ (red line), coupled MoS$_2$ (black line), and graphene layers (blue line) in MGHN-2, at $T_0$=300 K. Inset: a closer look for temperature distribution of the coupled graphene layer. (b) Schematic view of thermal transport process in MGHN-2. The horizontal and vertical arrows show heat fluxes in horizontal and vertical directions, respectively. The color bar labels the temperature scale in the structure.

**Figure 6.** (a) Thermal conductivity of MGHN-3 as a function of overlap ratio R$_{overlap}$, at $T_0$=300 K, $\eta$=1. (b) Thermal conductivities of MGHN-3 under different temperatures $T_0$ (100, 300, and 500 K) as functions of coupling strength $\eta$, at R$_{overlap}$ =50%.

**Figure 7.** (a) Horizontal temperature distributions of MoS$_2$ and graphene layers in MGHN-3 at $T_0$=300 K, with overlap ratio 50%. (b) Schematic view of thermal transport process in MGHN-3. The horizontal and vertical arrows show heat fluxes in horizontal and vertical directions, respectively. The color bar labels the temperature scale in the structure.

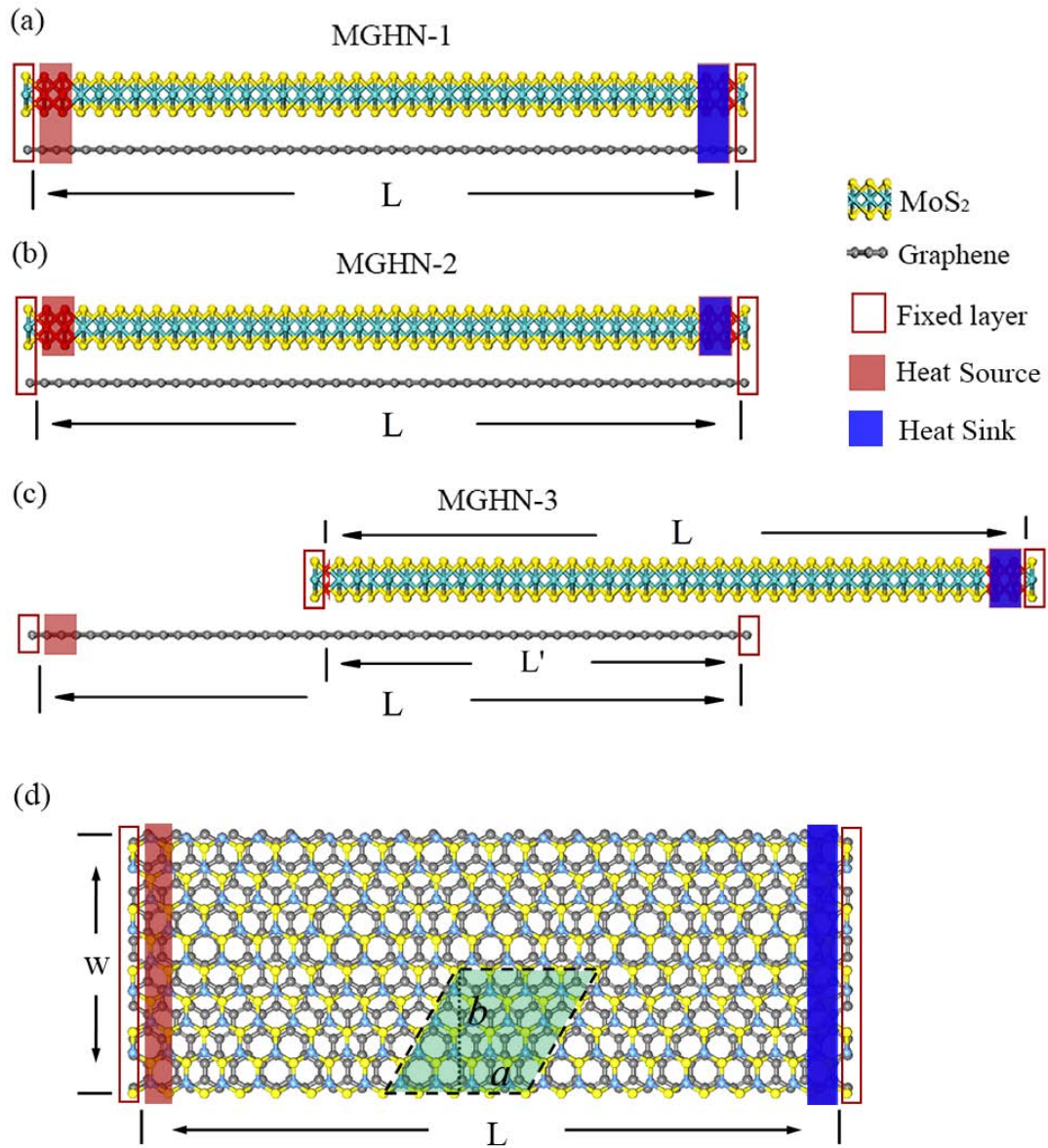

**Figure 1**

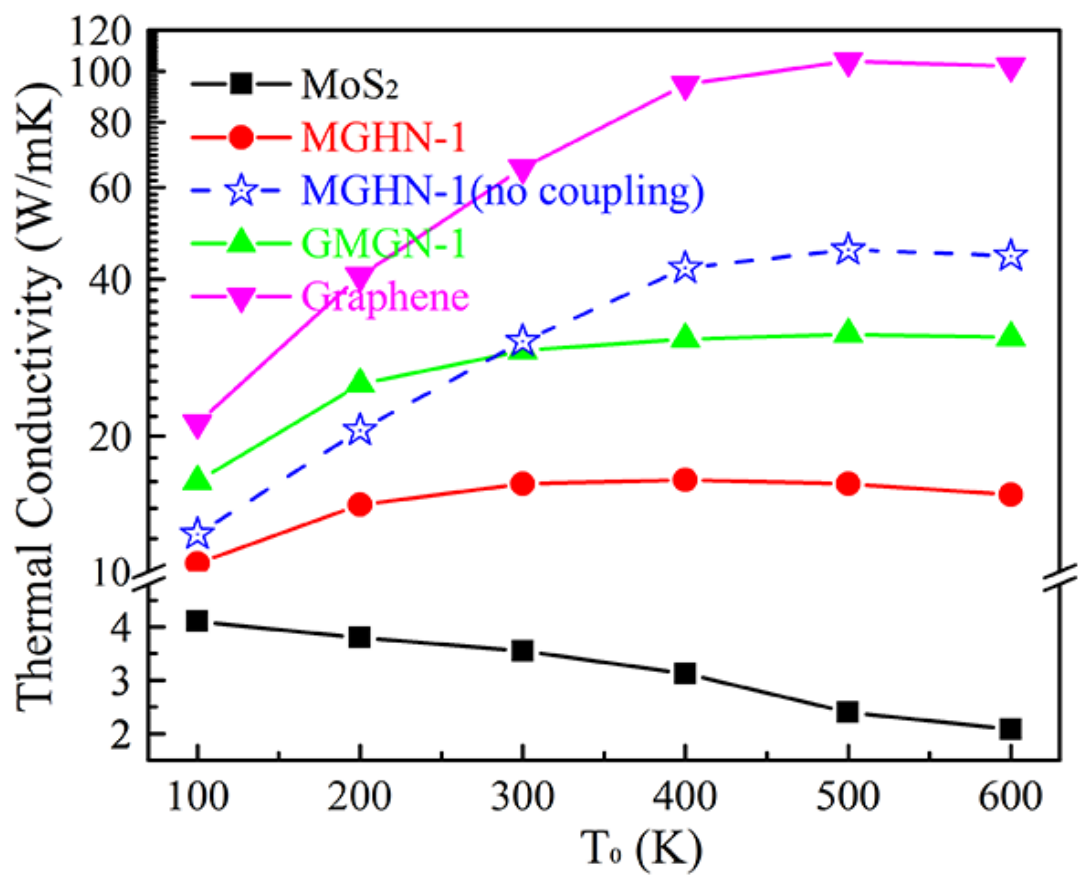

**Figure 2**

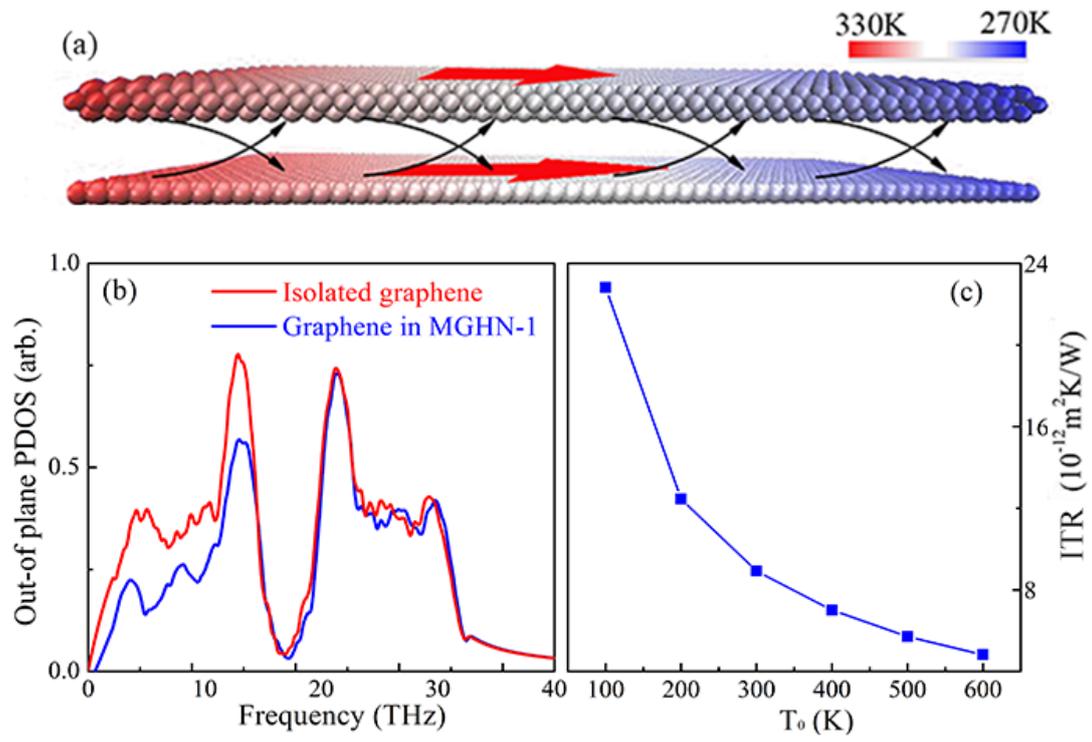

**Figure 3**

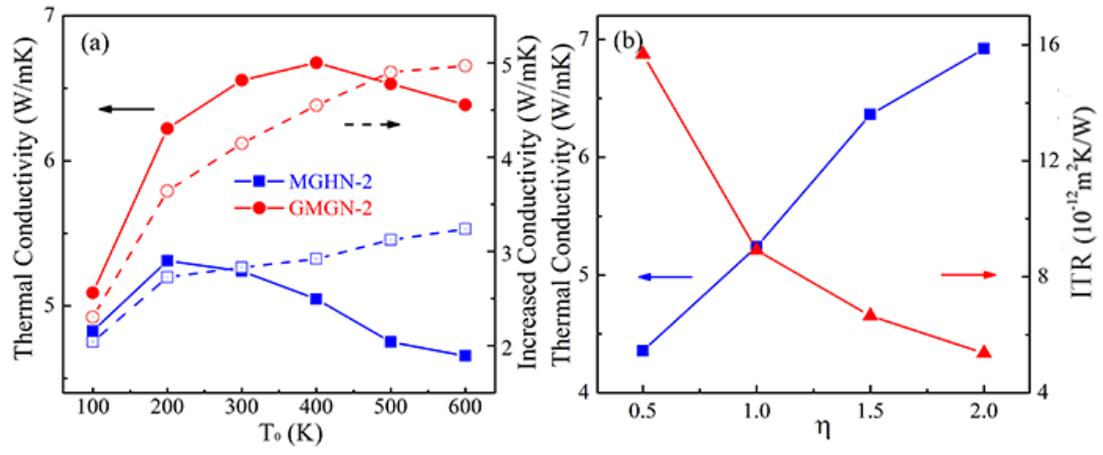

**Figure 4**

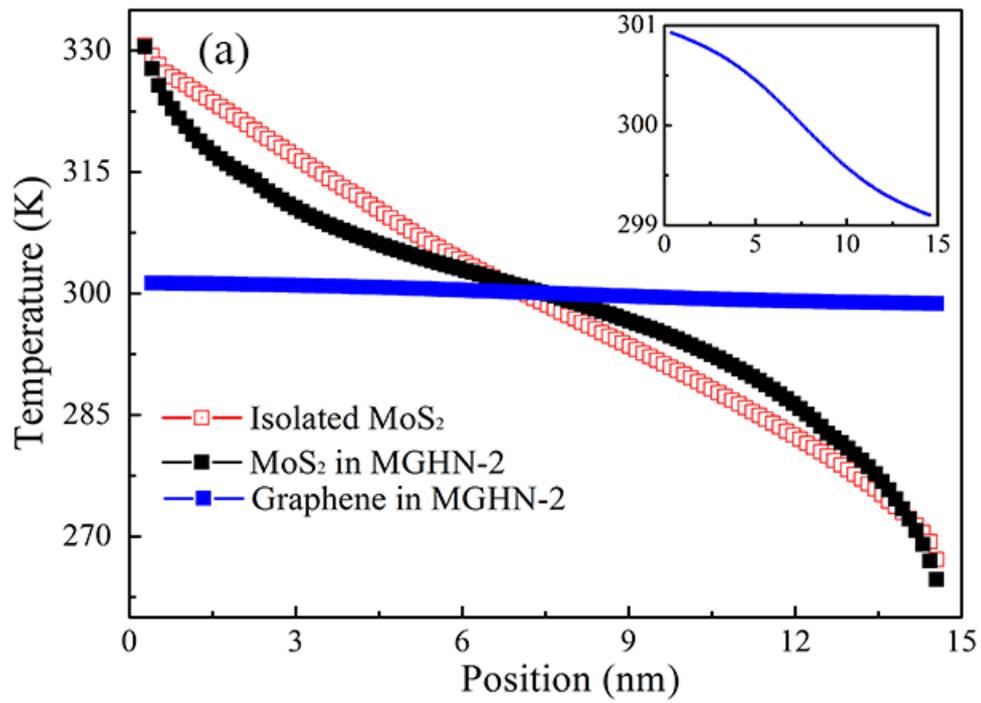

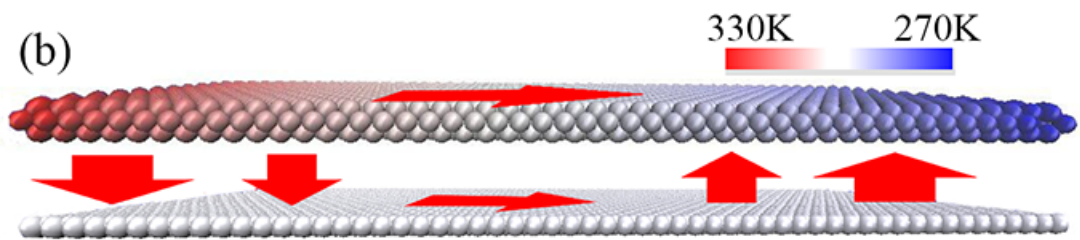

**Figure 5**

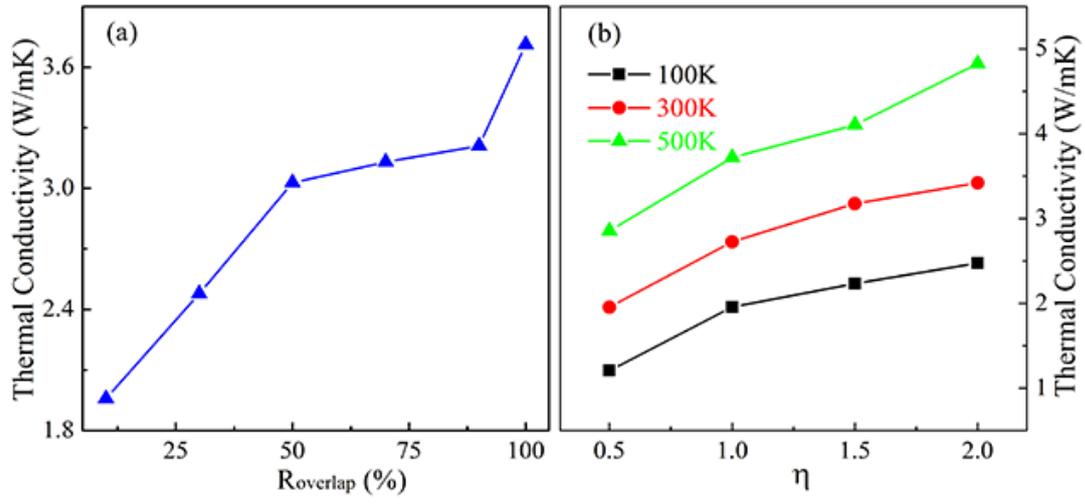

**Figure 6**

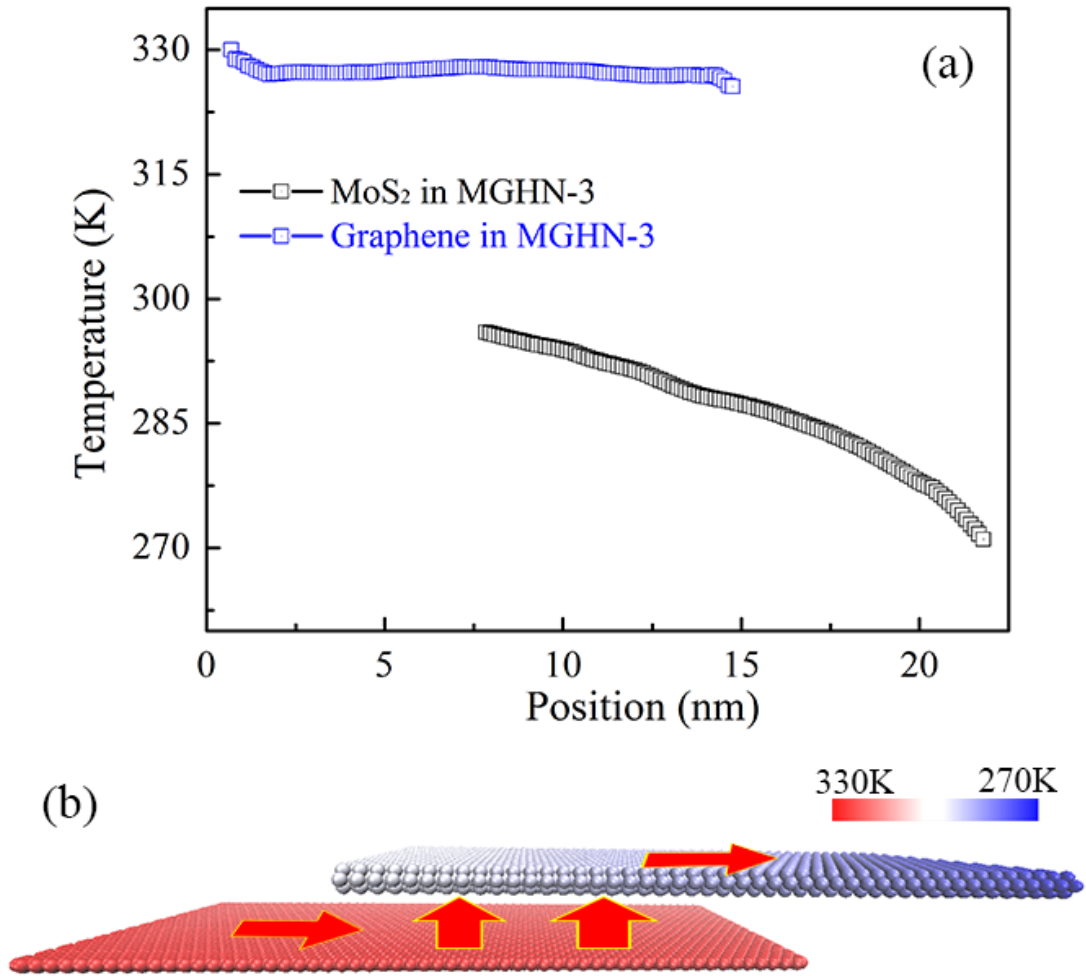

**Figure 7**